\documentclass[a4paper,11pt]{article}
\pdfoutput=1 % if your are submitting a pdflatex (i.e. if you have
\usepackage{jheppub} % for details on the use of the package, please
\usepackage{amsmath,epsf,amssymb,mathtools,latexsym,amsthm,setspace,array,pifont,hyperref,amsfonts,dsfont,cancel,braket,graphicx}

\usepackage{tensor}

\def\mL{\mathcal{L}}

\def\mO{\mathcal{O}}

\def\zb{\bar{z}}

\def\mb{\bar{m}}

\def\pa{\partial}

\renewcommand{\[}{\begin{equation}\begin{aligned}}
\renewcommand{\]}{\end{aligned}\end{equation}}

\def\g5{\gamma_5}

\def\Mt{\tilde{M}}

\def\Jt{\tilde{J}}

\def\Tt{\tilde{T}}

\def\b[#1]{\bold{#1}}
\def\bb[#1]{\overline{\bold{#1}}}
\def\bs[#1,#2]{\bold{#1}_{#2}}
\def\bbs[#1,#2]{\overline{\bold{#1}}_{#2}}

\def\s2{\sigma_2}
\def\ep{\epsilon}

\def\gammaflat{ \gamma_{z\zb}}

\def\paz{\pa_z}

\def\pazb{\pa_{\zb}}

\def\ketd[#1]{\ket{#1}_{\text{dressed}}}
\def\brad[#1]{\bra{#1}_{\text{dressed}}}
\def\ketas[#1]{\ket{#1}_{\text{Asymptotic}}}
\def\braas[#1]{\bra{#1}_{\text{Asymptotic}}}

\def\gammaflat{ \gamma_{z\zb}}

\def\paz{\pa_z}

\def\pazb{\pa_{\zb}}

\def\bp{\bold p}
\def\bA{\bold A}
\def\bB{\bold B}
\def\bD{\boldsymbol{\nabla} }

\def\bE{\bold E}
\def\bA{\bold A}
\def\bv{\bold v}

\def\bS{\bold S}

\def\eq{\begin{equation}}
\def\eqe{\end{equation}}
\def\eqa{\begin{eqnarray}}
\def\eqae{\end{eqnarray}}

\def\gc[#1,#2,#3]{\tensor{\Gamma}{_{#1#2}^{#3}}}
\def\torsion[#1,#2,#3]{\tensor{S}{_{#1#2}^{#3}}}
\def\contorsion[#1,#2,#3]{\tensor{K}{_{#1#2}^{#3}}}
\def\chris[#1,#2,#3]{\left\{\begin{array}{c}#1 \\#2#3 \end{array}\right\}}

\title{\boldmath Dual Komar Mass, Torsion and Riemann-Cartan Manifolds}
\author{Uri Kol}
\affiliation{Center for Cosmology and Particle Physics, Department of Physics, New York University, 726 Broadway, New York, NY 10003, USA}
\emailAdd{urikol@gmail.com}

\abstract{
The dual Komar mass generalizes the concept of the NUT parameter and is akin to the magnetic charge in electrodynamics.
In asymptotically flat spacetimes it coincides with the dual supertranslation charge.
The dual mass vanishes identically on Riemannian manifolds in General Relativity unless conical singularities corresponding to Misner strings are introduced.
In this paper we propose an alternative way to source the dual mass locally.
We show that this can be done by enlarging the phase space of the theory to allow for a violation of the algebraic Bianchi identity using local fields.
A minimal extension of Einstein's gravity that meets this requirement is known as the Einstein-Cartan theory.
Our main result is that on Riemann-Cartan manifolds the dual Komar mass does not vanish and is given by a volume integral over a local 1-form gravitational-magnetic current that is a function of the torsion.
}

\begin{document} 
\maketitle
\flushbottom

%%%%%%%%%%%%%%%%%%%%%%%%%%%%%%%%%%%%%%%%%%%%%%%%%%%%%%%%%%%%%%%%
%%%%%%%%%%%%%%%%%%%%%%%%%%%%%%%%%%%%%%%%%%%%%%%%%%%%%%%%%%%%%%%%
\section{Introduction}\label{sec:intro}
%%%%%%%%%%%%%%%%%%%%%%%%%%%%%%%%%%%%%%%%%%%%%%%%%%%%%%%%%%%%%%%%
%%%%%%%%%%%%%%%%%%%%%%%%%%%%%%%%%%%%%%%%%%%%%%%%%%%%%%%%%%%%%%%%

The Coulomb components of the electromagnetic field are characterized by their parity.
The electric component of the field is even under spatial reflections while its magnetic counterpart is odd.
Each one of these Coulomb field components gives rise to a conserved quantity, corresponding to the electric and magnetic charges.
Electric-magnetic duality rotates the field components and their associated charges into each other.
In gravity, the Coulomb components of the field can be classified according to their parity in a similar way and they give rise to two conserved quantities.
The mass is the conserved charge that is associated with the gravitational-electric component of the field, which is even under parity.
The gravitational-magnetic field component, which is odd under parity, gives rise to a second conserved charge known in the literature as the \emph{dual mass}, or the \emph{magnetic mass}, and is the main subject of this paper.

To motivate the discussion about the dual mass, let us start with a brief review of the Coulomb fields and their associated charges in electrodynamics and in gravity.
The Coulomb components of the electromagnetic field can be identified using the Newman-Penrose formalism \cite{Newman:1961qr,Janis:1965tx,Newman:1968uj}.
In this framework, the field strength $F_{\mu\nu}$ is organized in terms of three complex scalars
\begin{equation}
\begin{aligned}
\phi_0 &= F_{\mu\nu} \ell^{\mu}m^{\nu}, \\
\phi_1 &= \frac{1}{2}F_{\mu\nu} \left(\ell^{\mu}n^{\nu} - \mb^{\mu}m^{\nu}\right),\\
\phi_2 &= F_{\mu\nu} \mb^{\mu}n^{\nu},
\end{aligned}
\end{equation}
where $e^{\mu}_a=\{\ell^{\mu},n^{\mu},m^{\mu},\mb ^{\mu}\}$ is a basis of complex null tetrads in Minkowski background.
We now assume that the current densities are localized and that the space is empty outside the distribution of charges.
Under this assumption, the asymptotic behaviors of the complex scalars at large distances are given by
\begin{equation}
\phi_n = \frac{\phi_n^{(0)}}{r^{3-n}} +\dots .
\end{equation}
Spacetime can then be classified into three zones - near, intermediate and far - according to the different decay rates of the Newman-Penrose scalars.
$\phi_0$ is dominant in the near zone.
$\phi_1$ describes the Coulomb components of the field and is dominant in the intermediate zone.
$\phi_2$ describes the radiative components of the field and is dominant in the far zone.
Here we will focus on the Coulomb field, whose real and imaginary parts correspond, respectively, to its electric and magnetic components.
The associated conserved quantities are obtained by integrating the Coulomb fields over a large sphere
\begin{equation}
%\frac{1}{2\pi e}
\int_{S^2} d\Omega \, \phi_1^{(0)} = q+ i g,
\end{equation}
corresponding to the electric and magnetic charges, respectively.

The gravitational field can be expanded systematically in a similar way.
In the Newman-Penrose formalism \cite{Newman:1968uj}, the Weyl tensor $C_{\mu\nu\rho\sigma}$ is organized in terms of five complex scalars
\begin{equation}
\begin{aligned}
\psi_0 &= C_{\mu\nu\rho\sigma}
\ell^{\mu}m^{\nu}\ell^{\rho}m^{\sigma}
,\\
\psi_1 &= C_{\mu\nu\rho\sigma}
\ell^{\mu}n^{\nu}\ell^{\rho}m^{\sigma}
,\\
\psi_2 &=C_{\mu\nu\rho\sigma}
\mb^{\mu}n^{\nu}\ell^{\rho}m^{\sigma}
,\\
\psi_3 &= C_{\mu\nu\rho\sigma}
\mb^{\mu}n^{\nu}\ell^{\rho}n^{\sigma}
,\\
\psi_4 &=C_{\mu\nu\rho\sigma}
\mb^{\mu}n^{\nu}\mb^{\rho}n^{\sigma}.
\end{aligned}
\end{equation}
Assuming that the stress tensor is localized and that spacetime is empty outside the distribution of matter, the five complex scalars decay at large distances in the following way
\begin{equation}
\psi_n = \frac{\psi_n^{(0)}}{r^{5-n}} +\dots .
\end{equation}
The different decay rates now classify spacetime into five zones in which the different field components are dominant.
$\psi_0$ and $\psi_1$ describe near-fields. $\psi_2$ describes the Coulomb components of the field and is dominant in the intermediate zone.
$\psi_3$ and $\psi_4$ describe the radiative modes.
The real and imaginary parts of the Coulomb field correspond, respectively, to its gravitational-electric and gravitational-magnetic components.
The associated conserved quantities are obtained by integrating the Coulomb fields over a large sphere
\begin{equation}\label{massDef}
\frac{1}{4\pi G}\int_{S^2} d\Omega \, \psi_2^{(0)} = M+ i \Mt,
\end{equation}
corresponding to the mass $M$ and the dual mass $\Mt$, respectively.
The two conserved gravitational charges \eqref{massDef} are defined on a large sphere at spatial infinity, while the time coordinate is kept finite.
Here we are assuming that there is no gravitational radiation or any matter distribution at spatial infinity.

In asymptotically flat spacetimes $M$ will coincide with the Arnowitt, Deser and Misner (ADM) mass \cite{ADM} which is interpreted as the total energy.
However, the limit of a very large sphere can be taken in different ways.
For example, the Bondi mass \cite{Bondi:1960jsa,Bondi:1962px,Sachs:1962wk} is defined on a large sphere at \emph{null} infinity, where gravitational radiation is present in general.
In this limit the sphere is taken to be very large while the null coordinate is kept finite.
When defined at future null infinity, the Bondi mass describes the remaining energy at some retarded null time $u$ after the emission of gravitational radiation. Similarly one can define the Bondi mass at past null infinity, where it describes the energy stored in a given spacetime before the emission of gravitational radiation at some advanced null time $v$.
The Bondi mass therefore differs from the ADM mass by the total energy flux of the radiation that passed through the portion of null infinity to the past of $u$, or to the future of $v$.
When evaluated on the past boundary of future null infinity, or on the future boundary of past null infinity, the Bondi mass will therefore coincide with the ADM mass.

The purpose of the discussion above is to motivate, using the Coulomb components of the gravitational field, the study of the dual mass.
However, in the rest of the paper we will not use an asymptotic approximation nor limit our discussion to a particular asymptotic structure of spacetime.
Instead, we will use the Komar definition of mass \cite{Komar:1958wp}.
The Komar mass is defined solely in terms of a time-like Killing vector and does not rely on the asymptotic form of spacetime.
In his original paper, Komar considered spacetimes that obey an exact time-translation symmetry.
However, even if the time-like Killing vector is only approximate, the Komar definition of mass can still be used in those regions of spacetime where the symmetry holds to a good approximation.
For example, general asymptotically flat spacetimes obey an asymptotic time-translation symmetry and in those cases the Komar mass will coincide with the ADM and Bondi masses when evaluated at spatial and null infinities, respectively.
For more details we refer the reader to \cite{Wald:1984rg}.

There are two advantages for using the Komar formalism. The first, as already mentioned, is that it does not rely on a particular asymptotic structure of spacetime, in contrast to the ADM and Bondi masses.
However, the key feature of the Komar formalism is that it makes the analogy with electrodynamics manifest since the Komar mass is defined as the flux of a 2-form field strength, in a similar way to the electric charge.
This formulation therefore provides a natural definition of a \emph{dual mass} as the flux of the corresponding dual field strength, in analogy with the definition of the magnetic charge in electrodynamics. We will refer to the Komar mass and its dual collectively as the \emph{Komar charges}.

The Taub-NUT geometry is an example of a solution with a non-vanishing dual Komar mass, which is proportional in this case to the NUT parameter \cite{Bossard:2008sw}.
It exhibits a conical singularity akin to the Dirac string of a magnetic monopole.
The Taub-NUT geometry is \emph{locally} isomorphic to flat spacetime everywhere except at the location of the string.
However, \emph{globally} it has the topology of $\mathbb{R}\times S^3$ with the 3-sphere constructed as a non-trivial $S^1$ bundle over $S^2$.
The conical singularity can be eliminated using the Misner procedure \cite{Misner:1963fr} by dividing spacetime into two patches such that the metric is perfectly regular on each one of them separately. In the overlap region the solution is differentiable by virtue of diffeomorphisms at the cost of having to identify the time coordinate on the circle.
This procedure is mathematically identical to the construction of the Wu-Yang monopole in electrodynamics \cite{Wu:1976ge}.

In locally asymptotically flat spacetimes the dual Komar mass coincides with minus the dual supertranslation charge \cite{Godazgar:2018qpq,Godazgar:2019dkh}.
The later was shown to be a topological invariant on the space of asymptotic metrics \cite{Kol:2019nkc}. The dual mass therefore vanishes identically when the metric is globally defined. However, when the metric is not a globally defined 2-form, the dual mass can receive non-zero values. This can happen when a conical singularity is present, like in the Taub-NUT geometry, or when the metric is constructed as a section of a non-trivial bundle, like in the Misner space. It is therefore evident that the dual mass is directly related to the topological structure of spacetime.

As discussed above, the Taub-NUT metric exhibits a real physical conical singularity (as opposed to the Dirac string which is unobservable).
The attempt to eliminate the string using Misner's construction results in the appearance of the pathological closed timelike curves.
Recently, a new approach to the problem was introduced in \cite{Kol:2020ucd}, where the action of dual supertranslations on the asymptotic phase space was studied in detail.
Interestingly, it was found that this action does not correspond to spacetime diffeomorphisms. Instead, it was suggested in \cite{Kol:2020ucd} that dual supertranslation describe a redundant gauge symmetry since it acts trivially on S-Matrix elements.
These results provided a new way to construct the asymptotic form of a gravitational-magnetic monopole solution.
Instead of using diffeomorphisms to identify the two patches of the monopole bundle on the equatorial circle, the authors of \cite{Kol:2020ucd} employed the new dual supertranslation gauge symmetry. The advantage of this method is that it does not require the identification of the time coordinate on the circle.
It therefore provides an alternative to Misner's construction that avoids the pathological closed timelike curves.
It should be emphasized that this analysis was made in the asymptotic approximation and it remains to be seen whether it can be extended to the entire spacetime.

More generally, the dual Komar mass vanishes identically on any Riemannian manifold in General Relativity (GR) unless conical singularities corresponding to Misner strings are introduced.
Our main goal in this paper is to explore an alternative mechanism to source the dual mass \emph{locally}, without the need to consider extended wire singularities.
That can be done by sourcing, using local fields, the Bianchi identity of Komar's 2-form field strength (which is analogous to the electromagnetic field strength).
First we show that this Bianchi identity is closely related to the algebraic Bianchi identity of the Riemann tensor.
We are then led to consider an extension of the phase space in which the algebraic Bianchi identity can be sourced using local fields, as it is clearly impossible in GR \emph{per se}.
A minimal extension of Einstein's gravity that meets this requirement is known as the \emph{Einstein-Cartan} theory, which is formulated in the framework of \emph{Riemann-Cartan} geometry.
Riemann-Cartan manifolds, in turn, are equipped with an asymmetric affine connection, where its antisymmetric part is called the \emph{torsion}.
Our main result is that on Riemann-Cartan manifolds the dual Komar mass does not vanish and is given by a volume integral over a local 1-form gravitational-magnetic current that is a function of the torsion.
This mechanism is akin to the way in which magnetic monopoles are sources in electrodynamics.

Let us make a few comments about existing literature on the subject.
Similar ideas have appeared before in \cite{Henneaux:2004jw,Bunster:2006rt,Argurio:2009xr}, where a gravitational-magnetic source was introduced as a violation of the algebraic Bianchi identity in linearized gravity. However, the origin of this violation and its extension to the non-linear theory were not discussed there.
More recently, aspects of gravitational electric-magnetic duality between the standard and dual supertranslation charges were studied in \cite{Huang:2019cja} and the relation to the duality operation on curvature invariants is further explored in \cite{toAppear}.
The idea that the NUT charge can be interpreted in terms of torsion appeared in \cite{Bonnor_2001} but was not made precise and was mostly studied using mechanical examples.
Discussions about gravitational charges in the first order formalism can be found in \cite{Barnich:2016rwk,Barnich:2020ciy,Godazgar:2020gqd,Godazgar:2020kqd,Oliveri:2020xls}.

The paper is organized as follows. We start in section \ref{sec:DualKomar} with a review of the Komar formalism, focusing on the dual Komar mass, and discuss known examples of geometries with a non-zero dual mass. In section \ref{sec:Riemann} we set up the stage by analyzing the Komar charges on Reimannian manifolds. We emphasize that the Bianchi identity of Komar's 2-form field strength is directly related to the algebraic Bianchi identity of the Riemann tensor.
In section \ref{sec:Riemann-Cartan} we study the Komar charges on Riemann-Cartan manifolds. We show that the dual Komar mass can be sourced by a violation of the Bianchi identities discussed above and that the resulting gravitational-magnetic current is a function of the torsion. We also compute the contribution of the torsion to the gravitational-electric current.
In section \ref{sec:pointParticle} we study, as a physical application, the action of a point particle in the background of a Riemann-Cartan manifold and emphasize the roles of each one of the field components.
We end in section \ref{sec:Discussion} with a discussion and proposals for future research directions.
In appendix \ref{sec:appendix} we review the relationship between the Komar charges and the Bondi-Metzner-Sachs (BMS) charges in asymptotically flat spacetimes.

Throughout this work we will use the notations of \cite{Ortin:2004ms}.
More literature on torsion and Riemann-Cartan manifolds can be found in \cite{Wald:1984rg,Penrose:1987uia,Penrose:1986ca}.

%%%%%%%%%%%%%%%%%%%%%%%%%%%%%%%%%%%%%%%%%%%%%%%%%%%%%%%%%%%%%%%%
%%%%%%%%%%%%%%%%%%%%%%%%%%%%%%%%%%%%%%%%%%%%%%%%%%%%%%%%%%%%%%%%
\section{Dual Komar Mass}\label{sec:DualKomar}
%%%%%%%%%%%%%%%%%%%%%%%%%%%%%%%%%%%%%%%%%%%%%%%%%%%%%%%%%%%%%%%%
%%%%%%%%%%%%%%%%%%%%%%%%%%%%%%%%%%%%%%%%%%%%%%%%%%%%%%%%%%%%%%%%

In spacetimes that exhibit a time-like Killing vector $k=k^{\mu}\pa_{\mu}$ one can define a 1-form
\begin{equation}\label{oneForm}
A\equiv  g_{\mu\nu} k ^{\mu} dx^{\nu}
\end{equation}
and a corresponding field strength
\begin{equation}
F=dA.
\end{equation}
We will refer to $F$ as the \emph{Komar 2-form field strength}.
The Komar mass is then given by the surface integral over the Hodge dual of the field strength
\begin{equation}\label{KomarMass}
M= \frac{1}{8\pi G} \int _{\pa V} \star F,
\end{equation} 
where $\star$ is the Hodge star operation and $G$ is Newton's constant.
The Killing vector can be exact or approximate. In the first case the Komar integral can be defined, in principle, on any surface, while in the second case it will be meaningful only when evaluated on a surface in the region where the time-translation symmetry holds to a good approximation.
For example, asymptotically flat spacetimes obey an asymptotic time-translation symmetry which is part of the BMS group and therefore the Komar integral can be evaluated in the asymptotic region.

There are two advantages for using the Komar formalism. The first is that it does not rely on a linear approximation and in principle it can be applied in the non-linear theory and for every surface $\pa V$.
More importantly, it is formulated in a way that is similar to the definition of the electric charge in electrodynamics, where the mass is given by the flux of a 2-form field strength.
The analogy with electrodynamics motivates the definition of a dual Komar mass as the flux of the corresponding dual field strength
\begin{equation}\label{dualMass1}
\Mt = \frac{1}{8\pi G} \int _{\pa V}  F.
\end{equation} 
As in electrodynamics, when $\pa V$ is a closed surface and if $F$ is a globally defined 2-form, the total dual flux will vanish because it is given by an integral over a total derivative.

The field strength $F$ can be decomposed into "electric" and "magnetic" parts in the standard way
\begin{equation}\label{EMdecomposition}
F=  E_i dx ^i \wedge dx^0+B_i \star dx^i \wedge dx^0,
\end{equation}
where the index zero refers to the time coordinate and Latin indices run over the spatial coordinates. 
In terms of the 1-form potential we have
\begin{equation}\label{EMfields}
\bE = \bD A_0 -\nabla_0 \bA,  \qquad
\bB = \bD \times \bA,
\end{equation}
such that $A_0$ and $\bA$ play the roles of a scalar and vector potentials, respectively.
In regions where the time-translation symmetry is exact the time derivative term will vanish.
When evaluated on a spacelike hyper-surface, the Komar charges are then given by
\begin{equation}
\begin{aligned}
M    &= \frac{1}{8\pi G} \int _{\pa V} d \bS \cdot \bE, \\
\Mt &= \frac{1}{8\pi G} \int _{\pa V} d \bS \cdot \bB,
\end{aligned}
\end{equation}
correspond to the gravitational-electric and gravitational-magnetic fluxes, respectively.

The Komar charges can be converted in the standard way into volume integrals using the Stokes theorem
\begin{equation}\label{KomarMassVolume}
M= \frac{1}{8\pi G} \int _{V} d \star F, 
\end{equation}
\begin{equation}\label{dualMass2}
\Mt = \frac{1}{8\pi G} \int _{V}  dF.
\end{equation}
On a smooth Riemann surface the dual mass will therefore vanish due to the Bianchi identity
\begin{equation}
dF=d^2A =0
\end{equation}
of a 2-form field strength.
A violation of the Bianchi identity, however, can lead to a non-vanishing dual Komar mass, just as in electrodynamics.
That could happen when $A$ is not a globally defined 1-form.
In the following we will review two well-studied scenarios with a non-vanishing dual Komar mass.

\subsection{Taub-NUT Geometry}

The Taub-NUT metric is given by the following line element
\begin{equation}\label{TNmetric}
	ds^2=-f(r) \left(dt - 2 \ell \cos \theta d \varphi\right)^2 +\frac{dr^2}{f(r)} + \left(r^2 +\ell^2 \right) \left(d\theta ^2 +\sin^2 \theta d\varphi^2\right) ,
\end{equation}
with
\begin{equation}
	f(r) = \frac{r^2 -2mr - \ell^2}{r^2 +\ell ^2}.
\end{equation}
Here $m$ is the mass aspect and $\ell$ is called the NUT parameter. The Taub-NUT metric has two horizons located at $r_{\pm}=m \pm \sqrt{m^2 +\ell ^2}$. The Taub region is defined by the domain $r_-<r<r_+$ while the NUT region is defined by the domain $r>r_+$ and $r<r_-$.

The metric \eqref{TNmetric} is locally isomorphic to flat space everywhere except at $\theta=0$ and $\theta=\pi$ where there are conical singularities. The 1-form \eqref{oneForm} is given by
\begin{equation}
A= - f(r)dt +2 \ell f(r) \cos \theta d\varphi.
\end{equation}
Locally $dF=0$ everywhere except of $\theta=0,\pi$ where the metric is not differentiable. The dual Komar mass \eqref{dualMass1} on a large sphere is then given by
\begin{equation}\label{dualMassLarge}
\Mt = \lim_{r\rightarrow \infty} \frac{1}{8\pi G} \int _{S^2_r}  dA.
\end{equation} 
where $S^2_r$ is a two-sphere of radius $r$.
Here it does not matter how we take the large sphere limit since the geometry is static and there is no radiation.
Using Stokes theorem we can convert the surface integral into contour integrals around the conical defects in the north and south poles
\begin{equation}\label{TNmass}
	\Mt = 
	\lim_{r\rightarrow \infty} \frac{1}{8\pi G} \Big( \oint _{N}  A -\oint _{S}  A \Big) = \frac{\ell}{G}.
\end{equation} 
The relative minus sign between the two terms is due to the opposite orientation of the two contours.
We see that the dual Komar mass is proportional to the NUT parameter.
It is a straightforwards exercise to show that the Komar mass, evaluated in as similar way on a very large sphere, is given by
\begin{equation}
M=\frac{m}{G}.
\end{equation}
The Taub-NUT geometry is therefore akin to a dyon in electrodynamics, which has both electric and magnetic charges.

\subsection{Misner space}

The Taub-NUT metric contains a string-like singularity along the axes $\theta=0$ and $\theta=\pi$. It is possible to partly remove the singularity and shift it around using diffeomorphisms. For example, using the change of coordinates $t \rightarrow t + 2\ell \varphi$ we can remove the singularity at $\theta=0$ while the metric transforms into
\begin{equation}\label{North}
	ds_N^2=-f(r) \left(dt +4 \ell \sin^2 \frac{\theta}{2} d \varphi\right)^2 +\frac{dr^2}{f(r)} + \left(r^2 +\ell^2 \right) \left(d\theta ^2 +\sin^2 \theta d\varphi^2\right).
\end{equation}
The resulting metric contains a semi-infinite string singularity along the axis $\theta=\pi$. Similarly we can remove the singularity at $\theta=\pi$ by the change of coordinates $t \rightarrow t -2\ell \varphi$ applied to \eqref{TNmetric}. The resulting metric
\begin{equation}\label{South}
	ds_S^2=-f(r) \left(dt -4 \ell \cos^2 \frac{\theta}{2} d \varphi\right)^2 +\frac{dr^2}{f(r)} + \left(r^2 +\ell^2 \right) \left(d\theta ^2 +\sin^2 \theta d\varphi^2\right)
\end{equation}
contains a semi-infinite string singularity along the $\theta=0$ axis. It is not possible to completely remove the string singularity using diffeomorphisms, but only to change the direction of the semi-infinite string.

Misner \cite{Misner:1963fr} constructed a solution without string singularities by dividing the spacetime into two overlapping patches.
In the northern hemisphere he considered the metric \eqref{North} and in the southern hemisphere the metric \eqref{South}
\begin{equation}
ds^2= 
\begin{cases}
ds_N^2,    \qquad 0 \leq \theta  \leq \frac{\pi}{2} + \epsilon,
\\
ds_S^2, \qquad  \frac{\pi}{2} - \ep  \leq \theta \leq \pi .
\end{cases}
\end{equation}
The resulting geometry is perfectly regular in each one of the patches separately. To prove that the solution is also differentiable in the overlap region, Misner argued that in this region the two metrics are equivalent up to a diffeomorphism given by
\begin{equation}\label{diff}
	t_N = t_S -4 \ell \varphi.
\end{equation}
In other words, both the metric \emph{and} the spacetime manifold on which it is defined are divided into two patches in Misner's construction.
The consequence of \eqref{diff} is that since $\varphi$ is compact with a period of $2\pi$ then both $t_N$ and $t_S$ have to be compact with a period $8\pi \ell$ and the solution contains \emph{Closed Timelike Curves} (CTC).

The 1-form \eqref{oneForm} is now given by
\begin{equation}
A= 
\begin{cases}
A^N=- f(r)dt -4 \ell f(r) \sin^2 \frac{\theta}{2} d\varphi
,    \qquad 0 \leq \theta  \leq \frac{\pi}{2} + \epsilon,
\\
A^S= - f(r)dt +4 \ell f(r) \cos^2 \frac{\theta}{2} d\varphi
, \qquad  \frac{\pi}{2} - \ep  \leq \theta \leq \pi .
\end{cases}
\end{equation}
Locally $dF=0$.
However, note that $A$ is not globally defined anymore, but rather locally defined on the two patches separately.
The dual Komar mass on a large sphere \eqref{dualMassLarge} can be evaluated again using the Stokes theorem.
The surface integral is now converted into two contour integrals around the patches' boundaries, which are both given by the equatorial circle $S^1_r$
\begin{equation}
\Mt = 
\lim_{r\rightarrow \infty} \frac{1}{8\pi G}  \oint _{S^1_r}  \Big( A^N - A^S \Big) = \frac{\ell}{G}.
\end{equation} 
The relative minus sign between the two terms is, again, due to the opposite orientation of the two contours. 
The Misner space therefore admits a non-vanishing dual Komar mass, which is equal to that of the Taub-NUT metric \eqref{TNmass}.
Despite the fact that the Misner space is perfectly regular, we see that the dual Komar mass does not vanish because the metric is not globally defined.

%%%%%%%%%%%%%%%%%%%%%%%%%%%%%%%%%%%%%%%%%%%%%%%%%%%%%%%%%%%%%%%%
%%%%%%%%%%%%%%%%%%%%%%%%%%%%%%%%%%%%%%%%%%%%%%%%%%%%%%%%%%%%%%%%
\section{Riemannian Manifolds}\label{sec:Riemann}
%%%%%%%%%%%%%%%%%%%%%%%%%%%%%%%%%%%%%%%%%%%%%%%%%%%%%%%%%%%%%%%%
%%%%%%%%%%%%%%%%%%%%%%%%%%%%%%%%%%%%%%%%%%%%%%%%%%%%%%%%%%%%%%%%

As a warm up, we start by studying the Komar charges on Riemannian manifolds.
To see how the Komar mass \eqref{KomarMassVolume} is sourced, we first evaluate
\begin{equation}
\left(\star d \star F\right)_{\mu} = \nabla^{\nu}\left(dk\right)_{\nu\mu}= \nabla^{\nu} \nabla_{\nu} k_{\mu}-\nabla^{\nu} \nabla_{\mu} k_{\nu}.
\end{equation}
Now since $k^{\mu}$ is a Killing vector it obeys the Killing equation
\begin{equation}
\mL_k g_{\mu\nu} = -2 \nabla_{(\mu} k_{\nu)}= -2\left( \nabla_{\mu} k_{\nu}  +\nabla_{\nu} k_{\mu}      \right)=0
\end{equation}
and therefore
\begin{equation}\label{massCont}
\left(\star d \star F\right)_{\mu} =2 \nabla^{\nu} \nabla_{\nu} k_{\mu}=2 \nabla^{2} k_{\mu}=2 \nabla^{2} A_{\mu}.
\end{equation}
The Riemann tensor is defined using the commutator of two covariant derivatives acting, for example, on a covariant vector $V_{\rho}$
\begin{equation}\label{RiemannTensor}
\left[\nabla _{\mu}, \nabla_{\nu}\right] V_{\rho} = - V_{\sigma} \tensor{R}{_{\mu\nu\rho}^{\sigma}}.
\end{equation}
Applying the last equation to a Killing vector one can derive the following identity
\begin{equation}\label{KillingIdentity}
\nabla_{\nu}\nabla_{\rho} k_{\mu} = k_{\sigma}\tensor{R}{_{\rho \mu \nu}^{\sigma}}- \frac{1}{2}k_{\sigma}\tensor{R}{_{[\mu \nu \rho]}^{\sigma}}.
\end{equation}
where $\left[\dots\right]$ denote cyclic permutations.
The identity \eqref{KillingIdentity} is derived by taking cyclic permutations of \eqref{RiemannTensor} and using the Killing equation.
A Riemannian manifold obeys the \emph{algebraic Bianchi identity}
\begin{equation}\label{AlgebraicBianchi}
\tensor{R}{_{[\mu \nu \rho]}^{\sigma}}=0
\end{equation}
and therefore the second term on the right hand side of \eqref{KillingIdentity} vanishes. Finally, by contraction of indices one arrives at
\begin{equation}
\nabla^2 k_{\mu} = \tensor{R}{_{\mu}^{\nu}} k_{\nu},
\end{equation}
where $R_{\mu\nu}$ is the Ricci tensor. Therefore
\begin{equation}
\left(\star d \star F\right)_{\mu} =2 R_{\mu \nu} k^{\nu}=2 R_{\mu \nu} A^{\nu}.
\end{equation}
Using the Einstein equations
\begin{equation}\label{EinsteinTensor}
G_{\mu\nu} \equiv R_{\mu\nu} - \frac{1}{2}g_{\mu\nu} R = 8 \pi G T_{\mu\nu}
\end{equation}
we can therefore write
\begin{equation}
\left(\star d \star F\right)_{\mu} =16 \pi G \left(T_{\mu\nu} - \frac{1}{2}g_{\mu\nu} T \right) A^{\nu},
\end{equation}
where $T$ is the trace of the stress-energy tensor.

The 2-form $F$ therefore obeys the following set of equations
\begin{eqnarray}
\label{Maxwell} & d \star F &= \star J, \\
\label{Bianchi} & dF &= 0 ,
\end{eqnarray}
where the 1-form current is defined by
\begin{equation}\label{electricCurrent1}
J\equiv 16 \pi G \left(T_{\mu\nu} - \frac{1}{2}g_{\mu\nu} T \right) A^{\nu} dx^{\mu}.
\end{equation}
By equation \eqref{Maxwell} we see that $\star J$ is exact and conserved $d\star J =0 $.
Equation \eqref{Maxwell} takes the form of the inhomogeneous Maxwell equation, while \eqref{Bianchi} is the Bianchi identity of a 2-form field strength.
We therefore see that the Komar mass \eqref{KomarMassVolume} is given by the volume integral over the Hodge dual of the 1-form current \eqref{electricCurrent1}
\begin{equation}
M= \frac{1}{8\pi G} \int _{V} d \star F = 
\frac{1}{8\pi G} \int _{V} \star J
.
\end{equation}
On the contrary, the dual Komar mass is not sourced in Riemannian geometry and therefore vanishes identically on smooth manifolds.
These are standard results that can be found in the literature \cite{Wald:1984rg}.

\subsection{Bianchi Identities}

We have encountered two Bianchi identities so far; the algebraic Bianchi identity of the Riemann tensor \eqref{AlgebraicBianchi} and the Bianchi identity \eqref{Bianchi} of the 2-form field strength which is familiar from Maxwell's theory. Here we will show that these two Bianchi identities are related to each other.

The derivative of $F$ takes the following explicit form
\begin{equation}
\begin{aligned}
dF= d^2 A &=
\frac{1}{2!} \left(\nabla_{\rho} F_{\mu\nu}\right)\, dx^{\rho} \wedge dx^{\mu} \wedge dx^{\nu}
\\
&=\left( \nabla_{\rho}\nabla_{\mu}A_{\nu} \right)\, dx^{\rho} \wedge dx^{\mu} \wedge dx^{\nu}.
\end{aligned}
\end{equation}
This expression vanishes trivially when $A$ is a globally defined 1-form, like on a smooth Riemannian manifold.
However, let us press further and manipulate this Bianchi identity as follows
\begin{equation}
	dF=\frac{1}{2}\left( \left[\nabla_{\rho},\nabla_{\mu} \right] A_{\nu} \right)\, dx^{\rho} \wedge dx^{\mu} \wedge dx^{\nu},
\end{equation}
despite the fact that it is identically equal to zero.
Finally, by using the anti-symmetrization form we can write it as
\begin{equation}
dF=\frac{1}{2*3!}\left( \left[\nabla_{[\rho},\nabla_{\mu} \right] A_{\nu]} \right)\, dx^{\rho} \wedge dx^{\mu} \wedge dx^{\nu}.
\end{equation}
Now we can use the Ricci identity \eqref{RiemannTensor} to re-write the Bianchi identity of the 2-form field strength in terms of the Riemann tensor
\begin{equation}\label{BianchiRelation}
dF=- \frac{1}{2} \left( A_{\sigma}
\tensor{R}{_{[\rho \mu \nu]}^{\sigma}} \right)
\, dx^{\rho} \wedge dx^{\mu} \wedge dx^{\nu}.
\end{equation}
Equation \eqref{BianchiRelation} may seem trivial because on smooth Riemannian manifolds both sides vanish identically.
However, it does reveal the relation between the Bianchi identity of Komar's 2-form field strength and the algebraic Bianchi identity of the Riemann tensor.
It is therefore evident that in order to source the dual Komar mass one has to violate the algebraic Bianchi identity.

%%%%%%%%%%%%%%%%%%%%%%%%%%%%%%%%%%%%%%%%%%%%%%%%%%%%%%%%%%%%%%%%
%%%%%%%%%%%%%%%%%%%%%%%%%%%%%%%%%%%%%%%%%%%%%%%%%%%%%%%%%%%%%%%%
\section{Riemann-Cartan Manifolds}\label{sec:Riemann-Cartan}
%%%%%%%%%%%%%%%%%%%%%%%%%%%%%%%%%%%%%%%%%%%%%%%%%%%%%%%%%%%%%%%%
%%%%%%%%%%%%%%%%%%%%%%%%%%%%%%%%%%%%%%%%%%%%%%%%%%%%%%%%%%%%%%%%

In the previous section we have studied the Komar charges in Riemannian geometries.
On smooth Riemannian manifolds the dual Komar mass vanishes identically as a result of the Bianchi identity of the Komar field strength, which is directly related to the algebraic Bianchi identity of the Riemann tensor.
In this section we show how a violation of these Bianchi identities can lead to a non-vanishing dual mass.
Such a violation can occur on Riemann-Cartan manifolds in the presence of torsion and does not require the existence of a Misner string.

\subsection{Review}

Let us start with a brief review of the Riemann-Cartan geometry.
A Riemann-Cartan manifold is equipped with an affine connection that is not symmetric
\begin{equation}
\gc[\mu,\nu,\sigma]=\gc[(\mu,\nu),\sigma]+\gc[[\mu,\nu],\sigma].
\end{equation}
The symmetric part of the connection is $\gc[(\mu,\nu),\sigma]$, while its antisymmetric part is called the \emph{torsion} and it is a tensor
\begin{equation}
\torsion[\mu,\nu,\rho] \equiv -2 \gc[[\mu,\nu],\sigma] \equiv - \gc[\mu,\nu,\sigma]+\gc[\nu,\mu,\sigma] .
\end{equation}
The Riemann manifold is recovered when the torsion is set to zero.
The connection can be decomposed into its Riemannian and non-Riemannian parts as follows
\begin{equation}\label{connection}
\gc[\mu,\nu,\sigma] = \chris[\sigma,\mu,\nu]+
\contorsion[\mu,\nu,\sigma] ,
\end{equation}
where
\begin{equation}
\chris[\sigma,\mu,\nu]
=\frac{1}{2}g^{\sigma\rho}
\left(
\pa_{\mu} g_{\nu\rho} +\pa_{\nu} g_{\mu\rho} -\pa_{\rho} g_{\mu\nu} 
\right)
\end{equation}
are the  \emph{Christoffel symbols} and
\begin{equation}
\contorsion[\mu,\nu,\sigma]  = \frac{1}{2}g^{\sigma\rho}
\left(  S_{\mu\rho\nu}+S_{\nu\rho\mu} -S_{\mu\nu\rho} \right)
\end{equation}
is called the \emph{contorsion tensor}.
The contorsion obeys
\begin{equation}
\contorsion[[\mu,\nu],\sigma] = -\frac{1}{2}\torsion[\mu,\nu,\sigma],
\qquad
K_{\mu\nu\sigma} = - K_{\mu\sigma\nu}.
\end{equation}
Note that the metric and the torsion are independent degrees of freedom and that the contorsion depends on both.
Note also that the symmetric part of the connection receives a contribution from the contorsion and therefore depends on the torsion as well.

The Ricci identities for scalars and vectors in the presence of torsion take the form
\begin{equation}\label{RicciIdentities}
\begin{aligned}
\left[\nabla_{\mu} , \nabla_{\nu} \right] \phi & = + \torsion[\mu,\nu,\sigma] \nabla_{\sigma} \phi,\\
\left[\nabla_{\mu} , \nabla_{\nu} \right] V^{\rho} & =
+\tensor{R}{_{\mu \nu \sigma}^{\rho}}V^{\sigma}
+ \torsion[\mu,\nu,\sigma] \nabla_{\sigma} V^{\rho},\\
\left[\nabla_{\mu} , \nabla_{\nu} \right] V_{\rho} & =
-V_{\sigma }\tensor{R}{_{\mu \nu \rho}^{\sigma}}
+ \torsion[\mu,\nu,\sigma] \nabla_{\sigma} V_{\rho}.
\end{aligned}
\end{equation}
Note that the covariant derivatives are defined using the connection \eqref{connection} and therefore depend on the torsion.
Riemann-Cartan manifolds obey a set of Bianchi identities
\begin{eqnarray}
\tensor{R}{_{(\mu\nu) \rho}^{\sigma}} &=0, \\
\label{algBianchiTorsion}\tensor{R}{_{[\mu \nu \rho]}^{\sigma}} 
+\nabla_{[\mu}\torsion[\nu,\rho],\sigma]
+\torsion[[\mu,\nu,\alpha]\torsion[\rho],\alpha,\sigma]
&=0,\\
\label{thirdBianchi}\nabla_{[\alpha} \tensor{R}{_{\mu\nu]\rho}^{\sigma}}
+\torsion[[\alpha,\mu,\beta]\tensor{R}{_{\nu]\beta\rho}^{\sigma}}
&=0.
\end{eqnarray}
In particular, note that the algebraic Bianchi identity \eqref{AlgebraicBianchi} is now modified; on a Riemann-Cartan manifold it is sourced by the torsion tensor as in \eqref{algBianchiTorsion}.
Upon contraction of the $\nu$ and $\sigma$ indices in the third Bianchi identity \eqref{thirdBianchi}, one finds the \emph{contracted Bianchi identity}
\begin{equation}
\nabla_{\alpha} \tensor{G}{_{\mu}^{\alpha}} +2 S_{\mu\alpha\beta}R^{\beta\alpha}
-S_{\alpha\beta\nu}\tensor{R}{_{\mu}^{\nu\alpha\beta}}
=0
\end{equation}
that involves the Einstein tensor \eqref{EinsteinTensor}. A fourth Bianchi identity is given by
\begin{equation}
R_{\mu\nu(\sigma\rho)}=0.
\end{equation} 
However, note that unlike Riemannian geometry, in the presence of torsion there is no symmetry for exchanging the first two and last two indices of the Riemann tensor
\begin{equation}
R_{\mu\nu\sigma\rho} \neq R_{\sigma\rho \mu\nu } .
\end{equation}

\subsection{Dual Komar Mass}

As noted above, the algebraic Bianchi identity of the Riemann tensor is modified in the presence of torsion \eqref{algBianchiTorsion}.
Since the algebraic Bianchi identity is proportional to the Bianchi identity of the Komar field strength, its violation can potentially lead to a non-vanishing dual Komar mass.
Let us see how it works explicitly.

Having the machinery of the Riemann-Cartan geometry, we would now like to revisit the Bianchi identity of the Komar 2-form field strength \eqref{Bianchi} in the presence of torsion.
Given the modified algebraic Bianchi identity \eqref{algBianchiTorsion}, we can now repeat the steps that led to  \eqref{BianchiRelation} and compute the derivative of $F$ on a Riemann-Cartan manifold
\begin{equation}
\left(d F\right) _{\mu\nu\rho}=\frac{1}{2} \left[\nabla_{[\mu}, \nabla_{\nu}\right] A_{\rho]} 
=-\frac{1}{2}A_{\sigma }\tensor{R}{_{[\mu \nu \rho]}^{\sigma}}
- \frac{1}{2}\torsion[[\mu,\nu,\sigma] \nabla_{\rho]} A_{\sigma}.
\end{equation}
Here we have used that $\nabla_{\sigma} A_{\rho}=-\nabla_{\rho} A_{\sigma}$ (since $A_{\sigma}=k_{\sigma}$ is a Killing vector) and the Ricci identities in the presence of torsion \eqref{RicciIdentities}. Now using the algebraic Bianchi identity \eqref{algBianchiTorsion} we get
\begin{equation}
d F
=  \star \Jt = H ,
\end{equation}
where the dual, "magnetic", 1-form current
\begin{equation}
\Jt=\star H = \frac{1}{3!} \sqrt{\det g} \, \ep_{\mu\nu\sigma\rho} H^{\mu\nu\sigma} dx^{\rho}
\end{equation}
is given in terms of the 3-form $H$
\begin{equation}
H_{\mu\nu\rho}=\frac{1}{2}A_{\sigma }\left(
\nabla_{[\mu}\torsion[\nu,\rho],\sigma] +\torsion[[\mu,\nu,\alpha]\torsion[\rho],\alpha,\sigma]
\right)
- \frac{1}{2}\torsion[[\mu,\nu,\sigma] \nabla_{\rho]} A_{\sigma}.
\end{equation}
$\Jt$ is the gravitational-magnetic current.
In the presence of torsion we see that $dF$ does not vanish in general.
Therefore the dual Komar mass \eqref{dualMass2} is generally non-zero on a Riemann-Cartan manifold and given by a volume integral over a local charge density
\begin{equation}
\Mt = \frac{1}{8\pi G} \int _{V}  \star \Jt.
\end{equation}
This is our main result.

\subsection{Komar Mass}

The Komar mass also receives contributions from the torsion. This is not surprising, as it is well known that the torsion contributes to the stress-energy tensor. To compute the contribution of torsion to the mass we need to evaluate \eqref{massCont} on a Riemann-Cartan manifold. Using the Ricci identities \eqref{RicciIdentities} and the fact that $k$ is a Killing vector one can show that
\begin{equation}
\nabla_{\nu} \nabla_{\rho} k_{\mu} = 
k_{\sigma} \tensor{R}{_{\rho\mu\nu}^{\sigma}}
-\frac{1}{2}k_{\sigma}\tensor{R}{_{[\rho\mu\nu]}^{\sigma}}
- \torsion[\rho,\mu,\sigma] \nabla_{\sigma} k_{\nu}
-\frac{1}{2} \torsion[[\mu,\nu,\sigma] \nabla_{\rho]} k_{\sigma}.
\end{equation}
By contraction of indices in the equation above, we now have
\begin{equation}
\nabla^{2}k_{\mu}=\frac{1}{2}\left( 2\tensor{R}{_{\mu}^{\nu}}+\tensor{L}{_{\mu}^{\nu}}  +\tensor{\mO}{_{\mu}^{\nu}}  \right) k_{\nu} ,
\end{equation}
where
\begin{eqnarray}
\tensor{L}{_{\mu}^{\nu}} &\equiv& -g^{\sigma\rho}\tensor{R}{_{[\mu\sigma\rho]}^{\nu}}
=
g^{\sigma\rho}
 \left( 
+\nabla_{[\mu}\torsion[\sigma,\rho],\nu]
+\torsion[[\mu,\sigma,\alpha]\torsion[\rho],\alpha,\nu]
 \right)
, \\
\tensor{\mO}{_{\mu}^{\nu}}&\equiv& - g^{\sigma\rho} \left(
\torsion[[\mu,\sigma,\nu]\nabla_{\rho]}
-2 \torsion[\rho,\mu,\nu] \nabla_{\sigma}
\right) .
\end{eqnarray}
Note that $\tensor{\mO}{_{\mu}^{\nu}}$ is an operator. Both $\tensor{L}{_{\mu}^{\nu}} $ and $\tensor{\mO}{_{\mu}^{\nu}}$ vanish when the torsion is set to zero.

The "electric" component of the Komar 2-form $F$ then obeys
\begin{equation}
 d \star F = \star J
\end{equation}
where the gravitational-electric current is given by
\begin{equation}
J = \left( 2 \tensor{R}{_{\mu}^{\nu}}+ \tensor{L}{_{\mu}^{\nu}}  + \tensor{\mO}{_{\mu}^{\nu}}  \right) A_{\nu} dx^{\mu}.
\end{equation}
The Komar mass is therefore given by the integral over the Hodge dual of the gravitational-electric current
\begin{equation}
M = \frac{1}{8\pi G} \int _{V}  \star J = \frac{1}{8\pi G} \int _{V}  \star\left( 2 \tensor{R}{_{\mu}^{\nu}}+ \tensor{L}{_{\mu}^{\nu}}  + \tensor{\mO}{_{\mu}^{\nu}}  \right) A_{\nu} dx^{\mu}. 
\end{equation}

In Einstein's theory $\tensor{R}{_{\mu}^{\nu}}$ is proportional to the stress-energy tensor, but on a Riemann-Cartan manifold it will receive contributions from the torsion as well. The explicit equations of motion depend on the theory.
Here we will give an illustrative example, known as the Einstein-Cartan-Sciama-Kibble (ECSK) theory, which is the minimal extension of Einstein's theory that includes torsion \cite{Ortin:2004ms,Penrose:1986ca,Penrose:1987uia}.
Following the notations of \cite{Singh:2015ptw}, the ECSK action is given by
\begin{equation}
S= \int d^4x \sqrt{-g} \left(\frac{1}{16\pi G} R + \mL (\psi,\nabla\psi,g)\right).
\end{equation}
Here $\psi$ represent matter fields and the action depends on the torsion through the covariant derivative. The field equations are obtained by the variations of the action with respect to $\psi$, $g_{\mu\nu}$ and $K_{\mu\nu\sigma}$
\begin{equation}
\begin{aligned}
\frac{\delta \left(\sqrt{-g} \mL\right)}{\delta \psi} &=0 ,\\
\frac{\delta \left(\sqrt{-g} R\right)}{\delta g_{\mu\nu}} &=-16 \pi G \frac{\delta \left(\sqrt{-g} \mL\right)}{\delta g_{\mu\nu}}   ,\\
\frac{\delta \left(\sqrt{-g} R\right)}{\delta K_{\mu\nu\sigma}} &=-16 \pi G \frac{\delta \left(\sqrt{-g} \mL\right)}{\delta K_{\mu\nu\sigma}}   .
\end{aligned}
\end{equation}
The stress-energy tensor is defined in the usual way
\begin{equation}
T^{\mu\nu} = \frac{2}{\sqrt{-g}} \frac{\delta  \left(\sqrt{-g} \mL\right)}{ \delta g_{\mu\nu}}
\end{equation}
and the source of torsion is similarly defined by
\begin{equation}
\tau^{\mu\nu\sigma} =  \frac{2}{\sqrt{-g}} \frac{\delta  \left(\sqrt{-g} \mL\right)}{ \delta K_{\sigma\nu\mu}}.
\end{equation}
In many cases $\tau^{\mu\nu\sigma}$ can be identified with a spin density and therefore it is sometimes called the "spin angular momentum" or "spin-energy potential". However, more generally torsion does not necessarily describe spin. $\tau^{\mu\nu\sigma}$ is also referred to as the "hyper-momentum" in the literature.

The field equations of the ECSK theory are then given by
\begin{equation}\label{ECSKequations}
\begin{aligned}
G^{\mu\nu} &= 8\pi G \, \Sigma ^{\mu\nu}, \\
T ^{\mu\nu\sigma } &= 8 \pi G \, \tau^{\mu\nu\sigma},
\end{aligned}
\end{equation}
where $G^{\mu\nu} $ is the Einstein tensor,
\begin{equation}
\tensor{T}{_{\mu\nu}^{\sigma}} = \torsion[\mu,\nu,\sigma]+ \delta_{\mu}^{\sigma}\torsion[\nu,\lambda,\lambda]- \delta_{\nu}^{\sigma}\torsion[\mu,\lambda,\lambda]
\end{equation}
is called the \emph{modified torsion tensor} and
\begin{equation}
\Sigma^{\mu\nu} = T^{\mu\nu} +  \stackrel{\bigstar}{\nabla} _{\lambda}\left(\tau^{\mu\nu\lambda} - \tau^{\nu\lambda\mu} + \tau^{\lambda\mu\nu}\right)
\end{equation}
is the \emph{canonical stress-energy tensor}. Here $\stackrel{\bigstar}{\nabla} _{\lambda}\equiv \nabla_{\lambda}- \torsion[\lambda,\alpha,\alpha]$.
Note that the second equation in \eqref{ECSKequations} is an algebraic relation between the torsion and the hyper-momentum that can be solved explicitly.
One can therefore effectively cast the torsion out of the formalism by replacing it with the hyper-momentum.
Einstein's field equations are recovered when the torsion is set to zero.

The gravitational-electric current in the ECSK theory will then be given by
\begin{equation}
J= 
 \left( 16\pi G (\Sigma_{\mu\nu} - \frac{1}{2}g_{\mu\nu} \Sigma ) +  \tensor{L}{_{\mu\nu}}  +   \tensor{\mO}{_{\mu\nu}} \right)   A^{\nu} dx^{\mu},
\end{equation}
which now receives contributions from the torsion, both through the canonical stress-energy tensor $\Sigma_{\mu\nu}$ and through the terms proportional to $L_{\mu\nu}$ and $\mO_{\mu\nu}$.

%%%%%%%%%%%%%%%%%%%%%%%%%%%%%%%%%%%%%%%%%%%%%%%%%%%%%%%%%%%%%%%%
%%%%%%%%%%%%%%%%%%%%%%%%%%%%%%%%%%%%%%%%%%%%%%%%%%%%%%%%%%%%%%%%
\section{Point Particle}\label{sec:pointParticle}
%%%%%%%%%%%%%%%%%%%%%%%%%%%%%%%%%%%%%%%%%%%%%%%%%%%%%%%%%%%%%%%%
%%%%%%%%%%%%%%%%%%%%%%%%%%%%%%%%%%%%%%%%%%%%%%%%%%%%%%%%%%%%%%%%

In the previous sections we have studied the relations between the Komar charges, the torsion and the different metric components.
As an application of these results, we end here by studying the action of a point particle in the background of a Riemann-Cartan manifold using the Komar formalism.
This will serve as an example for the roles of the gravitational-electric and gravitational-magnetic field components in a physical scenario.

The action of a point particle is given by
\begin{equation}
S= \int d \tau  \, \sqrt{  g_{\mu\nu}  \dot{x}^{\mu} \dot{x}^{\nu}  },
\end{equation}
which essentially describes the length of a curve on the manifold.
The Euler-Lagrange equations of this action are the geodesic equations
\begin{equation}\label{geodesic}
\ddot{x}^{\mu} +\dot{x}^{\nu}\dot{x}^{\sigma} \chris[\mu,\nu,\sigma]=0.
\end{equation}
In other words, the geodesic equations minimize the action of a point particle. Note that in the presence of torsion these equations are different from the \emph{autoparallel equations}
\begin{equation}
\dot{x}^\nu \nabla_{\nu} \dot{x}^{\mu}=\ddot{x}^{\mu} +\dot{x}^{\nu}\dot{x}^{\sigma} \gc[\nu,\sigma,\mu]=0.
\end{equation}
that describe parallel transpose on Riemann-Cartan manifolds.

\subsection{Linear Approximation}

In the linear approximation, the only non-vanishing components of the Christoffel symbols that depend on $A$ are
\begin{equation}
\begin{aligned}
 \chris[0,i,j] &= - \frac{1}{2} \pa_{(i}A_{j)} , \qquad
&\chris[i,j,0] &= \frac{1}{2} \eta^{ik} \pa_{[j}A_{k]},\\
 \chris[0,0,i] &= - \frac{1}{2} \pa_i A_0 , \qquad
&\chris[i,0,0] &= -\frac{1}{2} \eta^{ij}\pa_j A_0.
\end{aligned}
\end{equation}
Here $\eta$ is the flat Minkowski metric, the zero index corresponds to the time coordinate and the indices $i,j$ run over the spatial coordinates.

Let us now consider a probe particle of mass $m_0$ moving in the linearized background. The geodesic equation \eqref{geodesic} of the probe particle in the linearized background then takes the form
\begin{equation}\label{geodesic2}
\frac{d \bp}{dt} = m_0 \gamma \left(
\frac{1}{2}\bE + \bv \times \bB
\right) +\dots ,
\end{equation}
where $\gamma$ is the Lorentz factor of the particle and $\bp$ its momentum.
The gravitational-electric $\bE$ and gravitational-magnetic $\bB$ fields are defined in \eqref{EMfields}.
The dots represent terms that depend on the $g_{ij}$ components of the metric.
These terms cannot be neglected, in general, but here we would like to focus on the contributions coming from the Komar 2-form and understand their physical meaning.

We can define
\begin{equation}
\phi \equiv \frac{1}{2} (A_0+1).
\end{equation}
In the Newtonian approximation $\phi$ will coincide with the Newtonian potential. For example, in the case of the Schwarzschild metric we have
\begin{equation}
\phi= \frac{m}{r}.
\end{equation}
We can now observe that the gravitational-electric force is equal to the gradient of $\phi$
\begin{equation}
\frac{1}{2}\bE  = \bD \phi.
\end{equation}
The factor of one half before the gravitational-electric force in \eqref{geodesic2} appears because we have defined the gravitational-electric field as a gradient of $A_0$ instead.

The effects of the gravitational-electric $\bE$ and gravitational-magnetic $\bB$ fields are now evident, as they induce on a point particle forces that are akin to the electric and magnetic fields in electrodynamics.

%%%%%%%%%%%%%%%%%%%%%%%%%%%%%%%%%%%%%%%%%%%%%%%%%%%%%%%%%%%%%%%%
%%%%%%%%%%%%%%%%%%%%%%%%%%%%%%%%%%%%%%%%%%%%%%%%%%%%%%%%%%%%%%%%
\section{Discussion}\label{sec:Discussion}
%%%%%%%%%%%%%%%%%%%%%%%%%%%%%%%%%%%%%%%%%%%%%%%%%%%%%%%%%%%%%%%%
%%%%%%%%%%%%%%%%%%%%%%%%%%%%%%%%%%%%%%%%%%%%%%%%%%%%%%%%%%%%%%%%

We have shown that on Riemann-Cartan manifolds, the Komar 2-form field strength obeys the following set of equations
\begin{equation}\label{MaxwellEqs}
\begin{aligned}
d \star F  &= \star J ,\\
dF &= \star \Jt .
\end{aligned}
\end{equation}
The gravitational-electric and gravitational-magnetic currents are functions of the metric and torsion, which are sourced in turn by the stress-energy tensor and by the hyper-momentum.
The set of equations \eqref{MaxwellEqs} takes the form of Maxwell's equations in electrodynamics with both electric and magnetic sources.
Though written in a simple form, however, note that these equations are non-linear in the fundamental fields, as opposed to Maxwell's equations of Abelian gauge theories.

The result \eqref{MaxwellEqs} implies that on a Riemann-Cartan manifold both the Komar 2-form and its dual are sourced \emph{locally}.
The Komar charges, defined as the fluxes of the 2-form $F$ and its dual, therefore do not vanish in Riemann-Cartan geometries.
In particular, our main result is that the dual mass can be sourced using local current densities, without the need to introduce extended conical singularities.

Riemannian geometry is recovered when the torsion is set to zero, in which case the dual current $\Jt$ vanishes. The Bianchi identity $dF=0$ then follows from the algebraic Bianchi identity of the Riemann tensor $\tensor{R}{_{[\mu \nu \rho]}^{\sigma}}=0$.
As a result, the dual mass vanishes, unless conical singularities that correspond to Misner strings are introduced.

As we have seen, the deviation from Riemannian geometry introduced by the manifold's torsion leads, in general, to a non-vanishing dual mass.
Further extensions of the theory can potentially lead to similar results.
In the framework of gravitational theories with an affine connection, known in the literature as Metric-Affine Gravity, one can also consider geometries with non-metricity.
However, non-metricity seems like a more radical deviation from Riemannian geometry.
The Riemann-Cartan manifold is therefore viewed as the minimal extension of the theory that leads to a violation of the algebraic Bianchi identity and as a consequence to a non-vanishing dual mass.

Our results suggest few interesting future directions to explore. First of all, we would like to find an explicit regular solution of a Riemann-Cartan manifold with a non-vanishing dual mass.
Another interesting direction would be to apply the Komar formalism for different Killing vectors and carry out a similar analysis for their corresponding conserved charges.
For example, in an axisymmetric spacetime there is a Killing vector that generates rotations about the axis of symmetry and which can be used to define angular momentum. It would be interesting to repeat our analysis for this scenario.

In this work we have focused on the equations that govern the dynamics of the Komar 2-form field strength.
We have carried out a local analysis and did not refer directly to the large distance behavior.
However, at the end of the day the Komar charges should be evaluated on a large surface and this is where the asymptotic form of spacetime comes into play.
Although we expect our formulas to hold in the general case, some subtleties arise when a cosmological constant is present (see for example \cite{Abbott:1981ff}).
Therefore, while the applications of our results to asymptotically flat spacetimes are straightforward, a closer attention should be paid when spacetime is asymptotically (Anti) de-Sitter and we plan to pursue this direction in the near future.

%%%%%%%%%%%%%%%%%%%%%%%%%%%%%%%%%%%%%%%%%%%%%%%%%%%%%%%%%%%%%%%%
%%%%%%%%%%%%%%%%%%%%%%%%%%%%%%%%%%%%%%%%%%%%%%%%%%%%%%%%%%%%%%%%
\acknowledgments
%%%%%%%%%%%%%%%%%%%%%%%%%%%%%%%%%%%%%%%%%%%%%%%%%%%%%%%%%%%%%%%%
%%%%%%%%%%%%%%%%%%%%%%%%%%%%%%%%%%%%%%%%%%%%%%%%%%%%%%%%%%%%%%%%

I would like to thank William Emond, Eliot Hijano, Yu-tin Huang, Reza Javadinezhad, Nathan Moynihan, Donal O'Connell and Massimo Porrati for useful discussions on related subjects.

%%%%%%%%%%%%%%%%%%%%%%%%%%%%%%%%%%%%%%%%%%%%%%%%%%%%%%%%%%%%%%%%
%%%%%%%%%%%%%%%%%%%%%%%%%%%%%%%%%%%%%%%%%%%%%%%%%%%%%%%%%%%%%%%%
\appendix
%%%%%%%%%%%%%%%%%%%%%%%%%%%%%%%%%%%%%%%%%%%%%%%%%%%%%%%%%%%%%%%%
%%%%%%%%%%%%%%%%%%%%%%%%%%%%%%%%%%%%%%%%%%%%%%%%%%%%%%%%%%%%%%%%

\section{BMS Charges}\label{sec:appendix}

In this appendix we will show that the Komar mass \eqref{KomarMass} and its dual \eqref{dualMass1} correspond to the standard and dual supertranslations charges in asymptotically flat spacetimes \cite{Kol:2019nkc,Godazgar:2018qpq,Godazgar:2019dkh,Barnich:2011mi,Strominger:2013jfa,He:2014laa}.

Asymptotically flat spacetimes can be expanded around future null infinity as follows
\begin{equation}\label{bondiMetric}
\begin{aligned}
ds^2 &= -du^2 -2dudr+2r^2 \gammaflat dz d\zb \\
&+
\frac{2m_B}{r} du^2 +r C_{zz} dz^2 +r C_{\zb\zb} d\zb ^2 -2U_z du dz -2 U_{\zb} du d\zb \\
&+ \dots .
\end{aligned}
\end{equation}
The first line above is the flat Minkowski metric, where the retarded null coordinate is
\begin{equation}
u=t-r
\end{equation}
and the complex coordinates on the two sphere are related to the standard angles by
\begin{equation}
z= \tan \frac{\theta}{2}e^{i\phi}.
\end{equation}
The metric on the unit two sphere is given by
\begin{equation}
 \gammaflat = \frac{2}{(1+z \zb)^2}.
\end{equation}
The second line in equation \eqref{bondiMetric} represent the leading correction to the flat metric at large distances and the dots represent subleading terms in the asymptotic expansion.
The radiative data $C_{zz}(u,z,\zb)$ is a function of all coordinates except $r$ and $U_z = - \frac{1}{2} D^z C_{zz}$.
The function $m_B(u,z,\zb)$ denotes the Bondi mass aspect.

The leading component of the Coulomb complex Weyl scalar is evaluated to be
\begin{equation}
\psi_2^{(0)} (u,z,\zb)= -m_B -\frac{1}{2}\left(\pa^z U_z -\pazb U_{\zb}\right) + \frac{1}{4} C^{zz}N_{zz}.
\end{equation}
The standard and dual supertranslation charges are defined at spatial infinity by
\begin{equation}
\begin{aligned}
T &= - \frac{1}{4\pi G} \int_{S^2} d\Omega \, \text{Re}  \, \psi_2^{(0)} = \frac{1}{4\pi G} \int_{S^2} d^2 z \gammaflat \,  m_B   , \\
\Tt &= - \frac{1}{4\pi G} \int_{S^2} d\Omega \, \text{Im} \,  \psi_2^{(0)} =\frac{i}{8\pi G} \int_{S^2} d^2z \, \left(\paz U_{\zb} - \pazb U_z\right),
\end{aligned}
\end{equation}
respectively.
Note that the Bondi news vanishes at spatial infinity
\begin{equation}
N_{zz} \Big| _{i^0} =N_{\zb\zb} \Big| _{i^0} = 0 ,
\end{equation}
per our assumption that there is no gravitational radiation there.

We now wish to evaluate the Komar charges on a very large sphere at spatial infinity, where we will assume that the metric is static. At asymptotic spatial infinity we therefore have
\begin{equation}
A_{\mu} = g_{t\mu}.
\end{equation}
Evaluating this on the metric \eqref{bondiMetric} we get
\begin{equation}
\begin{aligned}
A_0 dt  &= g_{00}dt = \left(-1 + \frac{2m_B}{r}\right) dt ,\\
A_i dx^i &= g_{0i} dx^i = - \frac{2m_B}{r}dr - U_z dz -U_{\zb}d\zb.
\end{aligned}
\end{equation}
The Komar mass can now be computed to give
\begin{equation}
M = \frac{1}{8\pi G} \int_{S^2} d^2z \gammaflat\, r^2 \, F_{rt} = - \frac{1}{4\pi G} \int_{S^2} d^2 z \gammaflat \,  m_B ,
\end{equation}
which is equal to minus the supertranslation charge $T$. The dual Komar mass in this case is
\begin{equation}
\Mt = \frac{i}{8\pi G} \int_{S^2} d^2z \, F_{z\zb}=-\frac{i}{8\pi G} \int_{S^2} d^2z \, \left(\paz U_{\zb} - \pazb U_z\right),
\end{equation}
coinciding with minus the dual supertranslation charge $\Tt$.

%%%%%%%%%%%%%%%%%%%%%%%%%%%%%%%%%%%%%%%%%%%%%%%%%%%%%%%%%%%%%%%%
%%%%%%%%%%%%%%%%%%%%%%%%%%%%%%%%%%%%%%%%%%%%%%%%%%%%%%%%%%%%%%%%
\bibliographystyle{JHEP}
\bibliography{bibliography}
%%%%%%%%%%%%%%%%%%%%%%%%%%%%%%%%%%%%%%%%%%%%%%%%%%%%%%%%%%%%%%%%
%%%%%%%%%%%%%%%%%%%%%%%%%%%%%%%%%%%%%%%%%%%%%%%%%%%%%%%%%%%%%%%%
\end{document}